\documentstyle[12pt,epsf]{article}
\begin{document}
\setcounter{page}{1}
\thispagestyle{empty}
\begin{flushright}
 JLAB-THY-98-39 \\
 hep-ph/9810254 \\
\end{flushright}
\vspace{5mm}
\begin{center}
\large
{\bf Isospin violation and the proton's neutral weak magnetic form factor} \\
\normalsize
\vspace{20mm}
Randy Lewis$^{1,2}$ and Nader Mobed$^2$ \\
{\it $^1$Jefferson Lab, 12000 Jefferson Avenue, Newport News, VA, U.S.A. 
         23606} \\
{\it $^2$Department of Physics, University of Regina, Regina, SK, Canada 
         S4S 0A2} \\
\vspace{10mm}
(October 1998) \\
\vspace{20mm}
Abstract \\
\end{center}
The effects of isospin violation on the neutral weak magnetic form factor of
the proton are studied using two-flavour chiral perturbation theory.
The first nonzero contributions appear at $O(p^4)$ in the small-momentum
expansion, and the $O(p^5)$ corrections are also calculated.
The leading contributions from an explicit $\Delta(1232)$ isomultiplet are 
included 
as well.  At such a high order in the chiral expansion, 
one might have expected a large number of unknown parameters to contribute.
However, it is found that no unknown parameters can appear within loop 
diagrams, and a single tree-level counterterm at $O(p^4)$ is sufficient to 
absorb all divergences.
The momentum dependence of the neutral weak magnetic form factor is 
not affected by this counterterm.

\newpage

\vspace{3mm}
\section{\small INTRODUCTION}

The first measurement of the proton's neutral weak magnetic form factor
has been reported recently by the SAMPLE Collaboration at 
MIT/Bates\cite{Bates}; subsequently, a series of precision experiments
has gotten underway at the Jefferson Lab\cite{JLab}.
Assuming only two quark flavours with exact isospin symmetry and neglecting
electroweak radiative corrections, the weak form factor can be expressed
in terms of the familiar electromagnetic form factors as follows:
\begin{equation}\label{basic}
   G_M^{p,Z}(q^2) = 
   \frac{1}{4}[G_M^p(q^2)-G_M^n(q^2)]-G_M^p(q^2){\rm sin}^2\theta_W~.
\end{equation}
Electroweak radiative corrections have been discussed in Ref. \cite{radcorr}.

There is presently a great deal of interest in determining the contribution
due to strange quarks\cite{physrep}, which simply appears as a new term
added to the right-hand side of Eq.~(\ref{basic}).
It is important to notice that isospin-violating effects also appear as a
new term on the right-hand side, even in the absence of strange quarks.
There have been some attempts to estimate the isospin-violating effects
by using constituent quark models\cite{DmiPol,CapRob,Miller} and a light-cone
meson-baryon fluctuation model\cite{Ma}.

In the present work, the effects of isospin violation are studied
using two-flavour heavy baryon chiral perturbation theory 
(HBChPT)\cite{JenMan,HBChPT}, which 
is nicely suited to the task.  At small momentum transfer, HBChPT is a 
systematic expansion in small parameters with dynamics arising from the 
propagation of pions and photons in the presence of a single baryon.
The remaining short-distance physics can only appear as low-energy constants
(i.e. parameters in the HBChPT Lagrangian).
The spontaneously-broken chiral symmetry of QCD is respected in HBChPT
by construction, and the explicit breaking due to current quark masses
can also be included in a systematic way.

Isospin violation occurs in nature through electromagnetic as well as strong 
interactions, so it is necessary to include the effects of virtual
photons in HBChPT.  The required Lagrangian has been constructed by
Meissner and Steininger\cite{emHBChPT}.

In principle, {\it three}-flavour HBChPT can be used to explicitly include
strangeness in the meson-cloud contribution to the neutral weak form factors,
and some work in this direction has recently been reported.\cite{su3}
However, the chiral expansion is not as well-behaved for strange quarks as
it is for up and down quarks,
and since the present work is only concerned with isospin-violating effects
it is preferable to work with two-flavour HBChPT.

The neutral weak vector form factors of the nucleons are defined here in the
notation of Dmitra\v{s}inovi\'{c} and Pollock\cite{DmiPol},
\begin{eqnarray}
\left<N(\vec{p}+\vec{q})|\frac{1}{2}(\bar{u}\gamma_\mu{u}-\bar{d}\gamma_\mu{d})
|N(\vec{p})\right> &=& \bar{u}(\vec{p}+\vec{q})\left[\frac{1}{2}(
{}^{u-d}F_1^{p+n}\pm{}^{u-d}F_1^{p-n})\gamma_\mu\right. \nonumber \\
             & & \left.+\frac{1}{2}({}^{u-d}F_2^{p+n}\pm
{}^{u-d}F_2^{p-n})\frac{i\sigma_{\mu\nu}q^\nu}{2M_N}\right]u(\vec{p}),
             \label{DmiPol1} \\
\left<N(\vec{p}+\vec{q})|\frac{1}{6}(\bar{u}\gamma_\mu{u}+\bar{d}\gamma_\mu{d})
|N(\vec{p})\right> &=& \bar{u}(\vec{p}+\vec{q})\left[\frac{1}{2}(
{}^{u+d}F_1^{p+n}\pm{}^{u+d}F_1^{p-n})\gamma_\mu\right. \nonumber \\
             & & \left.+\frac{1}{2}({}^{u+d}F_2^{p+n}\pm
{}^{u+d}F_2^{p-n})\frac{i\sigma_{\mu\nu}q^\nu}{2M_N}\right]u(\vec{p}),
\label{DmiPol2}
\end{eqnarray}
with $M_N$ denoting a nucleon mass.  
Walecka-Sachs form factors are defined by
\begin{eqnarray}
{}^iG_E^j(q^2) &=& {}^iF_1^j(q^2)+\frac{q^2}{4M_N^2}{}^iF_2^j(q^2), \\
{}^iG_M^j(q^2) &=& {}^iF_1^j(q^2)+{}^iF_2^j(q^2), \label{sach2}
\end{eqnarray}
where $i=u \pm d$ and $j=p \pm n$.
When isospin conservation is not enforced, Eq.~(\ref{basic}) generalizes to
\begin{equation}
G_M^{p,Z}(q^2) = \frac{1}{4}[G_M^p(q^2)-G_M^n(q^2)]-G_M^p(q^2){\rm sin}^2
   \theta_W-\frac{1}{4}G_M^{u,d}(q^2),
\end{equation}
where
\begin{equation}\label{GMud}
    G_M^{u,d}(q^2) \equiv {}^{u+d}G_M^{p-n}(q^2)-{}^{u-d}G_M^{p+n}(q^2)
\end{equation}
is the isospin-violating term.

In section \ref{preleading}, it is shown that isospin-violating
contributions do not appear up to and including $O(p^3)$ in HBChPT.
Section \ref{leading} presents and discusses the $O(p^4)$ calculation.
At this order,
the isospin-violating ${}^iF_1^j$ form factors contain no unknown parameters.
Each ${}^iF_2^j$ requires a single counterterm, but the momentum dependence
is not affected by the counterterm.

In section \ref{next2leading}, the calculation of $G_M^{u,d}(q^2)$
is extended to next-to-leading order, $O(p^5)$.
In principle, a large number
of low-energy constants could appear, some within loop diagrams and others
as tree-level counterterms, but it is shown
that the loop integrals are finite
and no new low-energy constants appear at this order.
The ratio of next-to-leading versus leading order contributions provides
some indication of the behaviour of the HBChPT expansion.
For the derivative of $G_M^{u,d}(q^2)$ at $q^2=0$, this ratio is close
to 1/2.

Section \ref{delta} evaluates and discusses the contributions to 
$G_M^{u,d}(q^2)$
that arise when the $\Delta(1232)$ is included explicitly in the chiral 
Lagrangian.
Section \ref{conc} offers a determination of the pion-cloud contribution
to the sole remaining parameter in $G_M^{u,d}(q^2)$, and then summarizes
the complete HBChPT result for $G_M^{u,d}(q^2)$.

\vspace{3mm}
\section{\small NO CONTRIBUTION UP TO $O(p^3)$}\label{preleading}

The Lagrangian of HBChPT is written in the form
\begin{equation}\label{LpiN}
   {\cal L}_{\pi{N}} = {\cal L}_{\pi{N}}^{(1)} + {\cal L}_{\pi{N}}^{(2)}
                     + {\cal L}_{\pi{N}}^{(3)} + {\cal L}_{\pi{N}}^{(4)}
                     + {\cal L}_{\pi{N}}^{(5)} + \ldots,
\end{equation}
where the superscripts denote powers in the ``momentum'' expansion,
which is actually a combined expansion in various small quantities.
A covariant derivative counts as one power (therefore the field strength
of an external current counts as two powers) and a current
quark mass counts as two powers (recall that $m_\pi^2 \sim m_q$).  
These dimensionful quantities are small relative to both the chiral scale
and the nucleon masses, $4\pi{F_\pi} \approx M_p \approx M_n$.
Eq.~(\ref{LpiN}) represents an expansion in the inverse nucleon masses
as well as the chiral expansion.
Furthermore, the effects of virtual photons can be
organized according to an expansion in the electromagnetic coupling, and
it is convenient to use $O(e) \sim O(p)$\cite{emHBChPT}, which allows the
virtual photon effects to also
be incorporated into the generalized momentum expansion of Eq.~(\ref{LpiN}).
In the present work, ``$O(p^n)$'' is used to denote $n$ powers of any of these
small quantities.

The lowest-order Lagrangian is
\begin{equation}\label{L1}
   {\cal L}_{\pi{N}}^{(1)} = \bar{N}_v(iv{\cdot}\nabla+g_AS{\cdot}u)N_v,
\end{equation}
where
\begin{eqnarray}
   N_v(x) &=& \exp\left[iM_0v{\cdot}x\right]\frac{1}{2}(1+v\!\!\!/)\psi(x), \\
   S_\mu &=& \frac{i}{2}\gamma_5\sigma_{\mu\nu}v^\nu, \\
   u_\mu &=& iu^\dagger(\partial_\mu-ir_\mu)u
            -iu(\partial_\mu-i\ell_\mu)u^\dagger, \\
   \nabla_\mu &=& \partial_\mu + \Gamma_\mu - iv_\mu^{(s)}, \\
   \Gamma_\mu &=& \frac{1}{2}\left[u^\dagger(\partial_\mu-ir_\mu)u
                  +u(\partial_\mu-i\ell_\mu)u^\dagger\right],
\end{eqnarray}
and $M_0$ is the lowest-order nucleon mass.
External vector and axial vector fields are included via 
$r_\mu = V_\mu + A_\mu$ and $\ell_\mu = V_\mu - A_\mu$,
and $u$ is a nonlinear representation of the pion fields, for example
\begin{equation}
   u = \exp\left[\frac{i}{2F}\left(\begin{tabular}{cc}
                 $\pi^0$ & $\sqrt{2}\pi^+$ \\
                 $\sqrt{2}\pi^-$ & $-\pi^0$ \end{tabular}\right)\right].
\end{equation}
The parameter $F$ corresponds to the pion decay constant in the chiral limit
(normalized according to $F_\pi \approx 93$ MeV).

The following relations will also be useful:
\begin{eqnarray}
S{\cdot}v &=& 0, \\
\{S_\mu,S_\nu\} &=& \frac{1}{2}(v_\mu{v}_\nu-g_{\mu\nu}), \\
\left[S_\mu,S_\nu\right] &=& i\epsilon_{\mu\nu\rho\omega}v^\rho{S}^\omega.
\end{eqnarray}
The relativistic currents required for this work can be re-expressed as
a $1/M_N$ expansion between HBChPT spinors.
In the rest frame of the initial nucleon, one finds
\begin{eqnarray}
\bar\psi(\vec{p}+\vec{q})\gamma_\mu\psi(\vec{p}) &=& 
   \bar{N}_v\left[v_\mu+\frac{q_\mu}{2M_N}
   +\frac{1}{M_N}i\epsilon_{\mu\nu\rho\sigma}q^\nu{v}^\rho{S}^\sigma
   +O\left(\frac{1}{M_N^2}\right)\right]N_v, \label{dirac1} \\
\bar\psi(\vec{p}+\vec{q})\frac{i\sigma_{\mu\nu}q^\nu}{2M_N}\psi(\vec{p}) &=& 
       \bar{N}_v\left[\frac{1}{M_N}i\epsilon_{\mu\nu\rho\omega}
       q^\nu{v}^\rho{S}^\omega + \frac{v_\mu{q^2}}{4M_N^2}
       + O\left(\frac{1}{M_N^3}\right)\right]N_v, \label{dirac2}
\end{eqnarray}
where $v_\mu = (1,0,0,0)$, $q_\mu$ is the 4-momentum of the incoming 
vector current and $M_N$ is the physical mass of the nucleon.

It is a simple matter to determine from ${\cal L}_{\pi{N}}^{(1)}$ the 
tree-level coupling of an external neutral vector field to a nucleon.  
The result is
\begin{eqnarray}
   {}^{u-d}F_1^{p-n} = {}^{u+d}F_1^{p+n} &=& 1, \label{tree1} \\
   {}^{u-d}F_1^{p+n} = {}^{u+d}F_1^{p-n} &=& {}^iF_2^j = 0,~{\rm for~all}~i,j.
   \label{tree2}
\end{eqnarray}
No contributions to the ${}^iF_2^j$ form factors can appear at this order
due to the explicit factor of $q/M_N$ in Eqs.~(\ref{DmiPol1}) and
(\ref{DmiPol2}).  Also, loop graphs are forbidden at $O(p)$ by
the standard power counting of HBChPT, which assures that one-loop diagrams
constructed from ${\cal L}_{\pi{N}}^{(1)}$ begin at $O(p^3)$.

The isospin-violating form factors must contain $m_u-m_d$ or a virtual photon
(with the associated factor of $e^2$), and are therefore suppressed by at least
two powers relative to the isospin-conserving ones.
This guarantees that ${}^{u-d}F_2^{p+n}$ and ${}^{u+d}F_2^{p-n}$ must remain
zero until $O(p^4)$.
One might expect ${}^{u-d}F_1^{p+n}$ and ${}^{u+d}F_1^{p-n}$ to
be nonzero at $O(p^3)$, but in fact they also remain zero until $O(p^4)$
due to Noether's theorem and the fact that 
the vector currents $\bar{u}\gamma_\mu{u}$ and $\bar{d}\gamma_\mu{d}$
are each conserved in QCD and QED (recall that weak radiative corrections 
are being neglected).  Noether's theorem requires that the ${}^iF_1^j$ 
form factors of Eqs.~(\ref{tree1}) and (\ref{tree2}) do not get 
renormalized at $q^2=0$, but
explicit factors of $q^2$ cannot appear until $O(p^4)$ in loop diagrams
or $O(p^5)$ at tree level.

Finally then, it is concluded that all isospin-violating form factors
remain zero up to and including $O(p^3)$.  It is instructive to verify this
fact by a direct calculation using the chiral Lagrangian which has been 
written down in its entirety at this order.\cite{EckM,Fettes,emHBChPT}
Following the notation of Ref. \cite{EckM} where field redefinitions
have been used to remove ``equation of motion'' terms from the Lagrangian,
the $O(p^2)$ terms which affect the neutral vector form factors are
\begin{equation}\label{L2bit}
   \delta{\cal L}_{\pi{N}}^{(2)} = \bar{N}_v\left[-\frac{\nabla\cdot\nabla}
   {2M_0}+\frac{1}{M_0}\epsilon^{\mu\nu\rho\sigma}v_\rho{S}_\sigma(
   a_6f_{+\mu\nu}+a_7v_{\mu\nu}^{(s)})\right]N_v,
\end{equation}
where
\begin{eqnarray}
   f_{+\mu\nu} &=& u\left(\partial_\mu\ell_\nu-\partial_\nu\ell_\mu
                   -i[\ell_\mu,\ell_\nu]\right)u^\dagger
                 + u^\dagger\left(\partial_\mu{r}_\nu-\partial_\nu{r}_\mu
                   -i[r_\mu,r_\nu]\right)u, \\
   v_{\mu\nu}^{(s)} &=& \partial_\mu{v}_\nu^{(s)}-\partial_\nu{v}_\mu^{(s)}.
\end{eqnarray}
Adding these terms to the lowest-order Lagrangian of Eq.~(\ref{L1}) leads
to the following coupling of an isoscalar vector current to a proton:
\begin{equation}
   \bar{N}_v\left[iv_\mu + \frac{i}{2M_0}(2k+q)_\mu 
   + \frac{2ia_7}{M_0}i\epsilon_{\mu\nu\rho\sigma}
   q^{\nu}v^\rho{S}^\sigma\right]N_v.
\end{equation}
According to Eq.~(\ref{L2bit}), 
the expression for an isovector current is obtained by the replacement
$a_7 \rightarrow 2a_6$.
$q_\mu$ is the incoming 4-momentum of the vector current, and 
the 4-vectors $v_\mu$ and $k_\mu$ are defined by
\begin{equation}
   p_\mu = M_0v_\mu + k_\mu,
\end{equation}
where $p_\mu$ is the 4-momentum of the incoming nucleon.

One is free to work in the rest frame of the initial nucleon, 
$p_\mu = (M_N,0,0,0)$, and to choose $v_\mu = (1,0,0,0)$.
For on-shell nucleons, these choices imply
\begin{eqnarray}
   k_\mu &=& (M_N-M_0)v_\mu \label{vk}, \\
   v{\cdot}q &=& \frac{-q^2}{2M_N},
\end{eqnarray}
and lead to 
\begin{eqnarray}\label{F1ord2}
   {}^{u+d}F_1^{p-n} = {}^{u-d}F_1^{p+n} &=& -\frac{(M_n-M_p)}{M_0}, \\
   {}^{u+d}F_2^{p-n} = {}^{u-d}F_2^{p+n} &=& 0.
\end{eqnarray}
Recalling that the neutron-proton mass difference is 
$O(1/M_0)$\cite{EckM,FLMS}, one
concludes that the nonzero result of Eq.~(\ref{F1ord2}) is suppressed by
two powers of $1/M_0$ relative to the leading isospin-conserving result.
In other words Eq.~(\ref{F1ord2}) contributes at
$O(p^3)$, and therefore the isospin-violating form factors all vanish
up to $O(p^2)$, as expected.

When the calculation is extended to $O(p^3)$, Eq.~(\ref{F1ord2}) is not
the only contribution.
${\cal L}_{\pi{N}}^{(3)}$, which includes virtual photons as well as strong
interactions, contains approximately 40 new 
parameters.\cite{EckM,Fettes,emHBChPT}  However, none of these parameters
can contribute to the isospin-violating form factors at $O(p^3)$, and the
only terms that do contribute are
\begin{equation}\label{L3}
   \delta{\cal L}_{\pi{N}}^{(3)} = \frac{1}{2M_0^2}\bar{N}_v\left(\left[
   \left(a_6-\frac{1}{8}\right)f_{+\mu\nu}+\left(a_7-\frac{1}{4}\right)
   v_{\mu\nu}^{(s)}\right]
   \epsilon^{\mu\nu\rho\sigma}S_\sigma{i}\nabla_\rho+{\rm h.c.}\right)N_v.
\end{equation}
A simple calculation shows that these terms do not affect ${}^iF_1^j$,
and their contributions to ${}^iF_2^j$ are effectively $O(p^4)$, 
so they are negligible in an $O(p^3)$ calculation.

In a general HBChPT calculation, one-loop diagrams built from 
${\cal L}_{\pi{N}}^{(1)}$ interactions can contribute at $O(p^3)$, 
but none
exist which can contribute to the isospin-violating form factors of interest
here.  However, there is a contribution from wave function 
re\-norm\-a\-liz\-a\-tion\cite{FLMS,EckM2}, 
\begin{equation}\label{Zord3}
   Z_p - Z_n = \frac{M_n-M_p}{M_0} + O\left(\frac{1}{M_0^3}\right),
\end{equation}
and it precisely cancels the $O(p^3)$ effect found in Eq.~(\ref{F1ord2}).
All $O(p^3)$ effects have now been discussed, so the isospin-violating
form factors do indeed remain zero at this order.

It is easy to verify that the final results of this section remain unchanged
if one does not employ the field redefinitions of Ref.~\cite{EckM}, even
though unphysical intermediate steps may differ.
For example, the wave function renormalization constant at $O(p^3)$ 
becomes independent of the nucleon mass when the field redefinitions are
not used, so $Z_p = Z_n$, but this change is exactly
compensated by the extra equation-of-motion term that would be present in
${\cal L}_{\pi{N}}^{(2)}$\cite{EckM},
\begin{equation}
   {\cal L}_{\pi{N}}^{(2)} \rightarrow {\cal L}_{\pi{N}}^{(2)}
   + \frac{1}{2M_0}\bar{N}_v(v\cdot\nabla)^2N_v.
\end{equation}

\vspace{3mm}
\section{\small LEADING ORDER, $O(p^4)$}\label{leading}

Having verified explicitly that the isospin-violating neutral weak vector
form factors
are exactly zero up to $O(p^3)$, the calculation will now be extended to
$O(p^4)$ where a nonzero result does exist.  The complete Lagrangian 
${\cal L}_{\pi{N}}^{(4)}$ has not been written in the literature
(but an effort is underway: the counterterms required for renormalization 
have recently been listed\cite{MMS}).  However, the contribution of
${\cal L}_{\pi{N}}^{(4)}$ to the present study is a simple constant and 
the full Lagrangian need not be constructed here.  Of greater interest are the
$q^2$-dependent $O(p^4)$ effects which come from loop diagrams
constructed using ${\cal L}_{\pi{N}}^{(1)}+{\cal L}_{\pi{N}}^{(2)}$.

\begin{figure}[tbh]
\epsfxsize=300pt \epsfbox[-75 200 475 650]{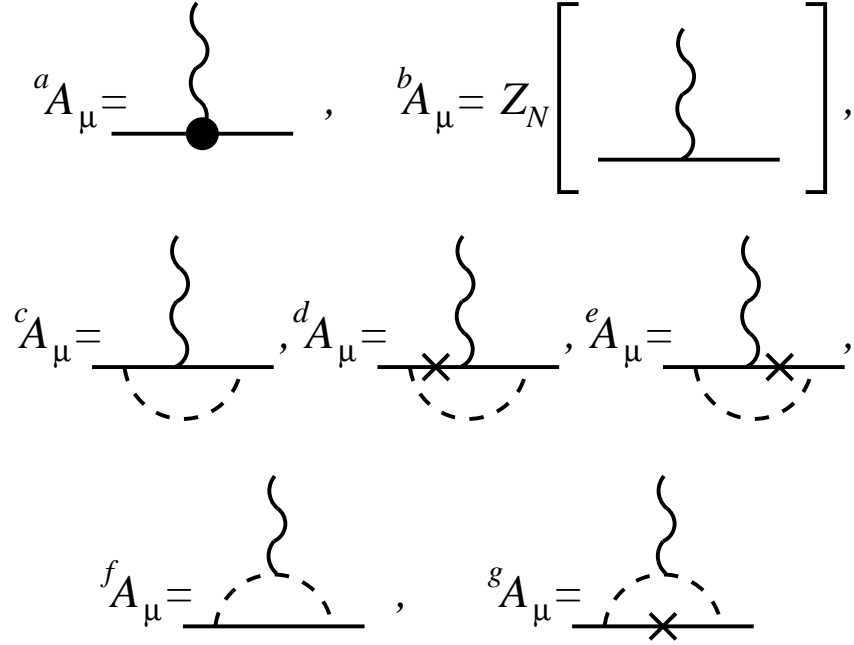}
\vspace{18pt}
\caption{\small Contributions to the isospin-violating
         vector form factors of a nucleon at $O(p^4)$.
         A dashed line represents the sum over charged and neutral pions.
         The solid dot denotes an insertion from ${\cal L}_{\pi{N}}^{(3)} +
         {\cal L}_{\pi{N}}^{(4)}$,
         a cross denotes an insertion from ${\cal L}_{\pi{N}}^{(2)}$,
         and all other Feynman rules come from ${\cal L}_{\pi{N}}^{(1)} +
         {\cal L}_{\pi\pi}^{(2)}$.
         }\label{O4diagrams}
\end{figure}

All of the pion-loop diagrams which contribute to the isospin-violating form 
factors at $O(p^4)$ are shown in Fig.~\ref{O4diagrams}.
(Diagrams with virtual photon loops will be discussed later in this section.)
Each of them will be evaluated in the rest frame of the initial nucleon
with the ``velocity'' parameter fixed to $v_\mu = (1,0,0,0)$.  Final
results for the form factors do not depend on these choices.

The pion propagator and the vector-pion couplings are obtained from the
lowest-order chiral Lagrangian for mesons,
\begin{eqnarray}
   {\cal L}_{\pi\pi}^{(2)} &=&
       \frac{F^2}{4}{\rm Tr}\left[D_\mu{U}^\dagger{D}^\mu{U}
       + 2B\left(\begin{array}{cc} m_u & 0 \\ 0 & m_d \end{array}
       \right)(U+U^\dagger)\right. \nonumber \\
    && \left.
       +e^2C\left(\begin{array}{cc} 2/3 & 0 \\ 0 & -1/3 \end{array}\right)U
       \left(\begin{array}{cc} 2/3 & 0 \\ 0 & -1/3 \end{array}\right)U^\dagger
       \right],
\end{eqnarray}
where $U \equiv u^2$ and
\begin{equation}
   D_\mu{U} = \partial_\mu{U} - ir_\mu{U} + iU\ell_\mu.
\end{equation}
Notice that the physical pion masses are nonzero for two reasons:
current quark mass effects (the parameter $B$ in ${\cal L}_{\pi\pi}^{(2)}$)
and electromagnetic effects (the parameter $C$ in ${\cal L}_{\pi\pi}^{(2)}$).

\begin{figure}[tbh]
\epsfxsize=300pt \epsfbox[-75 340 475 500]{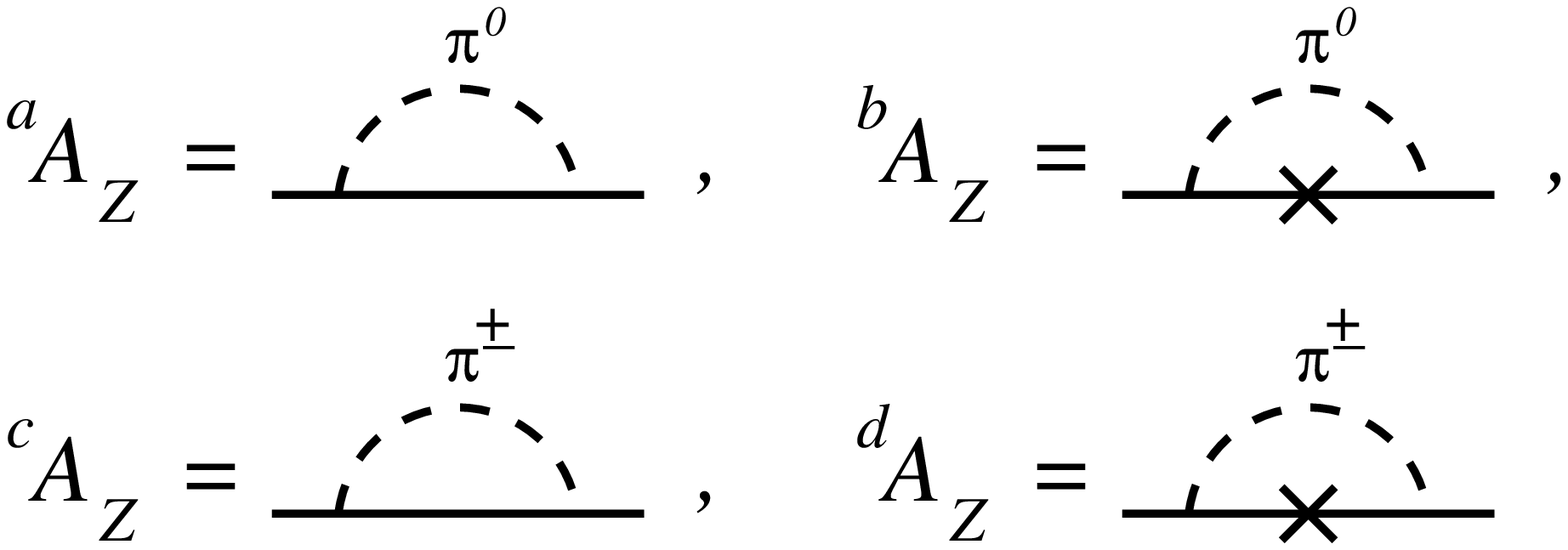}
\vspace{18pt}
\caption{\small 
         Contributions to the isospin-violating piece of the nucleon's wave
         function renormalization constant.
         A cross denotes an insertion from ${\cal L}_{\pi{N}}^{(2)}$,
         and all other Feynman rules come from ${\cal L}_{\pi{N}}^{(1)} +
         {\cal L}_{\pi\pi}^{(2)}$.
         }\label{Zdiagrams}
\end{figure}
The wave function renormalization constant for ${\cal L}_{\pi{N}}$
with the field redefinitions of Ref.~\cite{EckM}
has been determined previously up to $O(p^3)$\cite{FLMS,EckM2}, but the
present work will require an extension of the isospin-violating part to
$O(p^4)$.  The relevant diagrams are displayed in Fig.~\ref{Zdiagrams}.
Using dimensional regularization, the calculation of a diagram without 
an $O(p^2)$ insertion proceeds as follows:
\begin{eqnarray}
   {}^aA_Z &=& 
   \mu^{4-d}\int\frac{{\rm d}^d\ell}{(2\pi)^d}
   \left(\frac{i}{\ell^2-m_{\pi^0}^2+i\epsilon}\right)
   \left(-\frac{g_A}{F}S\cdot\ell\right)
   \left(\frac{i}{v\cdot(k-\ell)+i\epsilon}\right)\left(\frac{g_A}{F}S\cdot
   \ell\right) \\
   &=&
   \frac{-3ig_A^2}{4(4\pi{F})^2}\left[v{\cdot}k\left(m_{\pi^0}^2-\frac{2}{3}
   (v{\cdot}k)^2\right)\left(\frac{2}{4-d}+1-\gamma+{\rm ln}(4\pi)
   -{\rm ln}\frac{m_{\pi^0}^2}{\mu^2}\right)\right. \nonumber \\
 &&\left.+\frac{2}{3}v\cdot{k}\left(m_{\pi^0}^2-(v{\cdot}k)^2\right)
   -\frac{4}{3}\left(m_{\pi^0}^2-(v{\cdot}k)^2\right)^{3/2}\left(\frac{\pi}{2}
   +{\rm arcsin}\frac{v{\cdot}k}{m_{\pi^0}}\right)\right]. \label{Za}
\end{eqnarray}
Although this expression contains no explicit quark masses, it can still
violate isospin by virtue of the on-shell relation between $v{\cdot}k$ and the
nucleon masses as given in Eq.~(\ref{vk}).
The wave function renormalization constant is defined to be the residue of the
nucleon propagator at the on-shell point (multiplied by $i$), 
so Eq.~(\ref{Za}) should be viewed as a function of 
$x \equiv v{\cdot}k-M_N+M_0$.
The resulting contribution of this one diagram to the difference between
proton and neutron wave function renormalization constants is
\begin{equation}
   \delta(Z_p-Z_n) =
  i\frac{\rm d}{{\rm d}x}\left[{}^aA_Z(p)-{}^aA_Z(n)\right]_{x=0}
   = -\frac{3}{2}(M_n-M_p)\frac{\pi{g}_A^2m_{\pi^0}}{(4\pi{F})^2} 
   + O(M_n-M_p)^2.
\end{equation}
Similar contributions are made by the other diagrams in Fig.~\ref{Zdiagrams},
\begin{eqnarray}
  i\frac{\rm d}{{\rm d}x}\left[{}^bA_Z(p)-{}^bA_Z(n)\right]_{x=0}
  &=& \frac{3}{2}(M_n-M_p)\frac{\pi{g}_A^2m_{\pi^0}}{(4\pi{F})^2} 
   + O(M_n-M_p)^2, \\
  i\frac{\rm d}{{\rm d}x}\left[{}^cA_Z(p)-{}^cA_Z(n)\right]_{x=0}
  &=&-3(M_n-M_p)\frac{\pi{g}_A^2m_{\pi^+}}{(4\pi{F})^2} + O(M_n-M_p)^2, \\
  i\frac{\rm d}{{\rm d}x}\left[{}^dA_Z(p)-{}^dA_Z(n)\right]_{x=0}
  &=&-3(M_n-M_p)\frac{\pi{g}_A^2m_{\pi^+}}{(4\pi{F})^2} + O(M_n-M_p)^2.
\end{eqnarray}
Adding these four contributions
to the lower-order result of Eq.~(\ref{Zord3}) yields
\begin{equation}\label{deltaZ}
   Z_p - Z_n = \frac{M_n-M_p}{M_0}
           - 6(M_n-M_p)\frac{\pi{g}_A^2m_{\pi^+}}{(4\pi{F})^2}.
\end{equation}

The expressions for the matrix elements corresponding to the diagrams of
Fig.~\ref{O4diagrams} are now presented in the rest frame of the initial 
nucleon with $v_\mu = (1,0,0,0)$, on-shell external nucleons, and an
isoscalar vector current with incoming momentum $q_\mu$:
\begin{eqnarray}\label{ord4a}
   {}^aA^{(u+d)}_\mu(p)-{}^aA^{(u+d)}_\mu(n) &=& 
            {\rm const} \times i (M_n-M_p)
            i\epsilon_{\mu\nu\rho\sigma}q^\nu{v}^\rho{S}^\sigma, \\
   {}^bA^{(u+d)}_\mu(p)-{}^bA^{(u+d)}_\mu(n) &=& 
               -i(M_n-M_p)\left[6v_\mu\frac{\pi{g}_A^2m_{\pi^+}}{(4\pi{F})^2} 
               -\frac{q_\mu}{2M_0^2}
               -\frac{2a_7}{M_0^2}i\epsilon_{\mu\nu\rho\sigma}
               q^\nu{v}^\rho{S}^\sigma\right], \\
   {}^cA^{(u+d)}_\mu(p)-{}^cA^{(u+d)}_\mu(n) &=& 
            \frac{3}{2}iv_\mu(M_n-M_p)\frac{\pi{g}_A^2}{(4\pi{F})^2}
            (2m_{\pi^+}+m_{\pi^0}), \\
   {}^dA^{(u+d)}_\mu(p)-{}^dA^{(u+d)}_\mu(n) &=& 
   {}^eA^{(u+d)}_\mu(p)-{}^eA^{(u+d)}_\mu(n) \nonumber \\
          &=& \frac{3}{4}iv_\mu(M_n-M_p)\frac{\pi{g}_A^2}{(4\pi{F})^2}
            (2m_{\pi^+}-m_{\pi^0}), \\
   {}^fA^{(u+d)}_\mu(p)-{}^fA^{(u+d)}_\mu(n) &=& 
   {}^gA^{(u+d)}_\mu(p)-{}^gA^{(u+d)}_\mu(n) = 0. \label{ord4g}
\end{eqnarray}
Their sum is
\begin{equation}\label{totord4}
   A^{(u+d)}_\mu(p)-A^{(u+d)}_\mu(n) = 
               i(M_n-M_p)\left[ \frac{q_\mu}{2M_0^2}
               +{\rm const} \times i\epsilon_{\mu\nu\rho\sigma}
               q^\nu{v}^\rho{S}^\sigma\right]. \\
\end{equation}
The contribution from Eq.~(\ref{L3}) has also been absorbed into the
unspecified constant.
According to Eqs.~(\ref{dirac1}) and (\ref{dirac2}), the term containing
$i\epsilon_{\mu\nu\rho\sigma}q^\nu{v}^\rho{S}^\sigma$ in Eq.~(\ref{totord4})
is the leading contribution to ${}^{u+d}F_2^{p-n}(q^2)$.  If there had been a
term containing $v_\mu$, it would have been the leading contribution to
${}^{u+d}F_1^{p-n}(q^2)$.
As discussed in the previous section, ${}^{u+d}F_1^{p-n}(0)=0$ is 
required by Noether's theorem, and this is explicitly verified by the 
exact cancellation of terms proportional to $v_\mu$ in 
Eqs.~(\ref{ord4a})-(\ref{ord4g}).

There is also a term in Eq.~(\ref{totord4}) that is proportional to $q_\mu$,
and its origin is more subtle.  Instead of contributing to the isospin
violation of the form factors, this $q_\mu$ term is a consequence of
the nucleon mass dependence in the currents themselves.
Recall that Eqs.~(\ref{dirac1}) and (\ref{dirac2}) are expressed in terms
of the physical nucleon mass rather than the bare mass.
Thus,
\begin{equation}\label{qterm}
   \left[\bar\psi(\vec{p})\gamma_\mu\psi(\vec{p})\right]_p -
   \left[\bar\psi(\vec{p})\gamma_\mu\psi(\vec{p})\right]_n = 
   \frac{(M_n-M_p)}{2M_0^2}\bar{N}_v\left[q_\mu+2i\epsilon_{\mu\nu\rho\sigma}
   q^\nu{v}^\rho{S}^\sigma+O\left(\frac{1}{M_0^4}\right)\right]{N}_v.
\end{equation}
This term containing $q_\mu$ is precisely the one shown in Eq.~(\ref{totord4}),
while the term containing the antisymmetric tensor is absorbed into the
unspecified constant in Eq.~(\ref{totord4}).
Therefore, the $O(p^4)$ expressions for the ``$u+d$'' vector form factors are
\begin{eqnarray}
   {}^{u+d}F_1^{p-n}(q^2) &=& 0, \\
   {}^{u+d}F_2^{p-n}(q^2) &=& {\rm const} \equiv {}^{u+d}\kappa^{p-n}.
   \label{F2sing}
\end{eqnarray}

The form factors for an isovector (``$u-d$'') vector current
are less trivial because they receive a 
nonvanishing contribution from those diagrams in Fig.~\ref{O4diagrams}
where the vector current couples directly to a pion.  Those are the diagrams
which produce momentum dependence at $O(p^4)$.
The complete set of ``$u-d$'' contributions leads to
\begin{eqnarray}\label{F1trip}
   {}^{u-d}F_1^{p+n}(q^2) &=& -\frac{12\pi{g}_A^2}{(4\pi{F})^2}m_{\pi^+}
         (M_n-M_p)\left[1-\frac{4}{3}\int_0^1{\rm d}x\,\sqrt{1-x(1-x)\frac
         {q^2}{m_{\pi^+}^2}}\right. \nonumber \\
      && \left.+\frac{1}{3}\int_0^1{\rm d}x\left(1-x(1-x)\frac{q^2}
         {m_{\pi^+}^2}\right)^{-1/2}\right], \\
   {}^{u-d}F_2^{p+n}(q^2) &=& {}^{u-d}\kappa^{p+n}
         +\frac{16g_A^2M_N}{(4\pi{F})^2}(M_n-M_p)\int_0^1{\rm d}x\,{\rm ln}
         \left(1-x(1-x)\frac{q^2}{m_{\pi^+}^2}\right). \label{F2trip}
\end{eqnarray}
The numerical value of ${}^{u-d}\kappa^{p+n}$ is not
specified by chiral symmetry.
It should be noted that Eq.~(\ref{F1trip}) satisfies the constraint that 
${}^{u-d}F_1^{p+n}(0) = 0$, which is a
nontrivial check of the algebra since it results from a cancellation among the
diagrams of Fig.~\ref{O4diagrams}.

Throughout this section, pion-loop diagrams have been evaluated
while the effects of virtual photon loops were tacitly omitted.
This omission can be justified by simple power-counting arguments.
To begin, recall that the chiral Lagrangian is order-by-order renormalizable,
and that any two-loop diagram cannot contribute before $O(p^5)$.
(Photon loops have the same power counting as pion loops due to the
association $O(e) \sim O(p)$\cite{emHBChPT}.)
Up to $O(p^4)$ then, the only effects of virtual photons come from the
addition of a single virtual photon to the simplest tree-level form factor 
diagram.

Such a one-photon-loop diagram could offer an $O(p^3)$ contribution to
${}^iF_1^j(q^2)$.  However, this form factor is required to vanish at
$q^2=0$, so there must be an extra factor of $q^2/M_0^2$ at least, which then
contributes at $O(p^5)$.  This stands in contrast to the pion-loop diagrams
which contain the extra mass scale $m_\pi$; the ratio $q^2/m_\pi^2$ does
not lead to extra suppression in the HBChPT expansion.

A one-photon-loop diagram could also offer a contribution to
${}^iF_2{}^j(q^2)$, which begins at $O(p^4)$ rather than $O(p^3)$,
due to the explicit factor of $q/M_N$ in the definition of these form factors,
Eqs.~(\ref{DmiPol1}) and (\ref{DmiPol2}).  The contribution must be a
simple constant because any $q^2$ dependence would require extra factors
of $1/M_0$ and would thus contribute at a higher order.
The constant $O(p^4)$ contribution from virtual photons simply adjusts the
numerical value of the unspecified parameters, ${}^i\kappa^j$, in
Eqs.~(\ref{F2sing}) and (\ref{F2trip}).

\begin{figure}[tbh]
\epsfxsize=380pt \epsfbox[30 419 498 732]{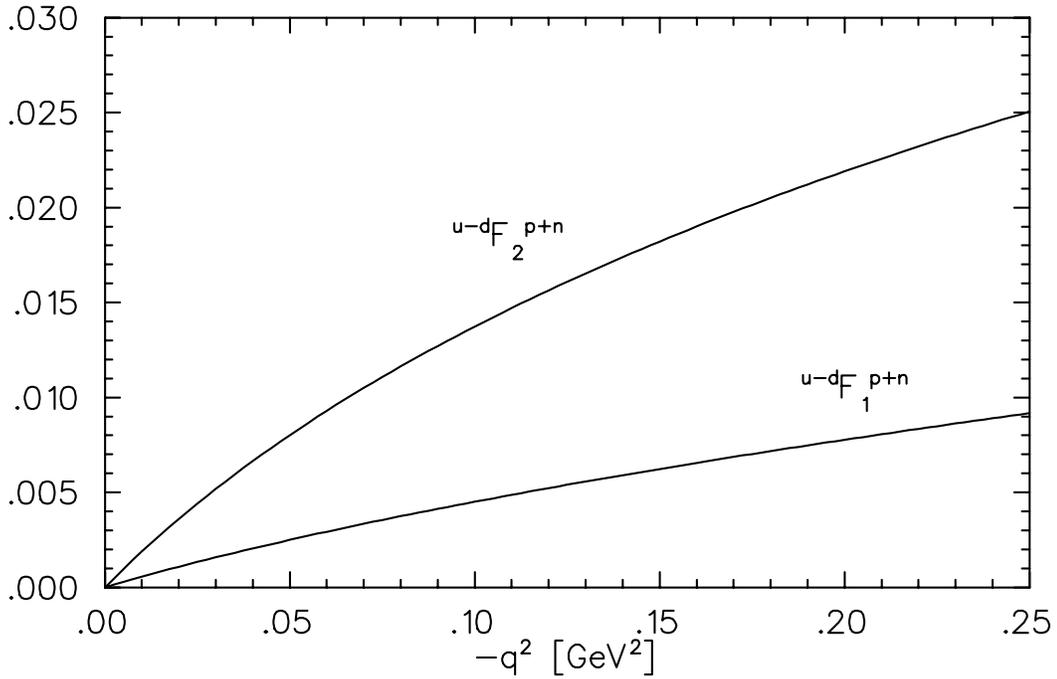}
\vspace{18pt}
\caption{\small
         Leading-order HBChPT results for the isospin-violating isovector
         form factors.  For this plot, the unspecified constant in 
         ${}^{u-d}F_2^{p+n}$
         has been chosen such that the form factor vanishes at $q^2=0$.
         }\label{O4plot}
\end{figure}
The $O(p^4)$ isospin-violating ``$u-d$'' form factors are plotted in 
Fig.~\ref{O4plot}
for $0 < -q^2 < 0.25~{\rm GeV}^2$, with ${}^{u-d}\kappa^{p+n}$ set to zero
and the numerical values of all other quantities set to those of the Particle
Data Group.\cite{PDG}  Uncertainties are not shown in Fig.~\ref{O4plot}.
As will be discussed in the following section, the dominant uncertainty 
comes from the truncation of the HBChPT expansion at $O(p^4)$, and an
estimate of this uncertainty will be obtained from the $O(p^5)$ calculation.

Independent of the value of ${}^{u-d}\kappa^{p+n}$,
Fig.~\ref{O4plot} indicates that both ``$u-d$'' form factors are
monotonically increasing at leading order, with ${}^{u-d}F_2^{p+n}$ 
increasing more quickly than ${}^{u-d}F_1^{p+n}$.

According to Eqs.~(\ref{sach2})-(\ref{GMud}), 
it is the sum of these two form factors which is
relevant to the proton's neutral weak magnetic form factor.
However, ${}^iF_1^j$ and ${}^iF_2^j$ differ by an explicit power of $M_N$
due to their definition, Eqs.~(\ref{DmiPol1}) and (\ref{DmiPol2}).
Thus, the leading order result from HBChPT is
\begin{equation}
   G_M^{u,d}(q^2) = {}^{u+d}\kappa^{p-n} - {}^{u-d}\kappa^{p+n}
             - \frac{16g_A^2M_N}{(4\pi{F})^2}(M_n-M_p)
               \int_0^1{\rm d}x\,{\rm ln}\left(1-x(1-x)\frac{q^2}{m_{\pi^+}^2}
               \right).
\end{equation}
The $O(p^4)$ result for ${}^{u-d}F_1^{p+n}$ is a next-to-leading-order 
correction to $G_M^{u,d}$, and it is reassuring to see from Fig.~\ref{O4plot}
that its $q^2$-dependence is smaller than the leading $q^2$-dependence.

The remaining next-to-leading-order effects come from an $O(p^5)$
calculation of ${}^iF_2^j$, and are discussed in the following section.

\vspace{3mm}
\section{\small NEXT-TO-LEADING ORDER, $O(p^5)$}\label{next2leading}

The next-to-leading corrections for both the ${}^iF_1^j$ and ${}^iF_2^j$ 
isospin-violating form factors occur at $O(p^5)$ in HBChPT.
The goal of the present section is to complete the next-to-leading
calculation of $G_M^{u,d}$.
${}^iF_1^j$ is itself a subleading contribution to $G_M^{u,d}$, so the
$O(p^4)$ result of the previous section is a next-to-leading order effect.
Therefore only ${}^iF_2^j$ needs to be calculated at $O(p^5)$.

To construct the set of contributing Feynman diagrams, consider first
those diagrams which contain no propagating pions (i.e. tree-level diagrams
or diagrams containing any number of photon loops).
The only available dimensionful parameters are the nucleon mass and
momentum transfer, and it is easily concluded that such diagrams cannot
contribute to the isospin-violating ${}^iF_2^j$ at $O(p^5)$.
Also, no contribution emerges from any diagram which contains both
one photon loop and one pion loop.

A diagram containing two pion loops can only contribute at $O(p^5)$ if
all Feynman rules come from ${\cal L}_{\pi{N}}^{(1)}+{\cal L}_{\pi\pi}^{(2)}$.
However, no isospin violation is contained within ${\cal L}_{\pi{N}}^{(1)}$,
and the $m_{\pi^+}-m_{\pi^0}$ mass difference does not affect the form
factors of interest here, so only one-pion-loop diagrams can contribute.

\begin{figure}[tbh]
\epsfxsize=300pt \epsfbox[-75 280 475 550]{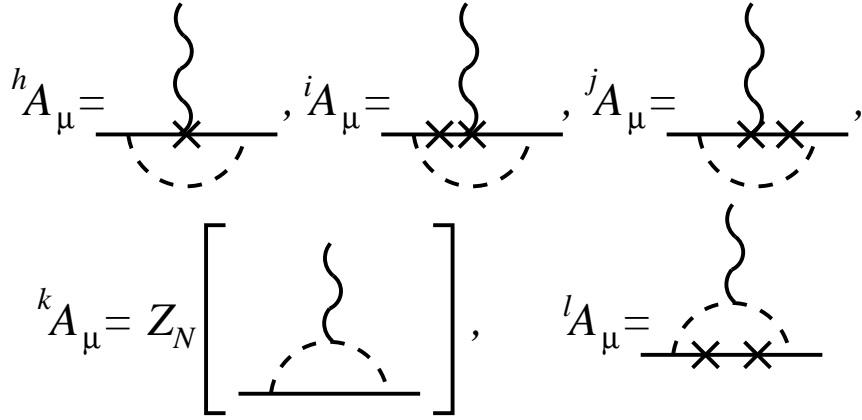}
\vspace{18pt}
\caption{\small
         Contributions to the isospin-violating ${}^iF_2^j$ form factors that
         begin at $O(p^5)$.
         A dashed line represents the sum over charged and neutral pions.
         A cross denotes an insertion from ${\cal L}_{\pi{N}}^{(2)}$,
         and all other Feynman rules come from ${\cal L}_{\pi{N}}^{(1)} +
         {\cal L}_{\pi\pi}^{(2)}$.
         }\label{O5diagrams}
\end{figure}
The set of diagrams which do contribute includes ${}^bA_\mu$,
${}^fA_\mu$ and ${}^gA_\mu$ in Fig.~\ref{O4diagrams}, plus
the diagrams of Fig.~\ref{O5diagrams}.
Although ${\cal L}_{\pi{N}}^{(3)}$ contains about 40 parameters that are 
unconstrained by chiral symmetry and could in principle appear within
$O(p^5)$ loop diagrams, none of them contribute to this calculation.  
Dimensional arguments do not permit a tree-level $O(p^5)$ counterterm for
${}^iF_2^j$,
(essentially because the small expansion parameters without uncontracted 
Lorentz indices tend to come in pairs at tree-level, such as $m_\pi^2$, $e^2$,
or $q^2$, but $O(p^5)$ would require an odd power),
so the total loop calculation must be finite.

Summing the $O(p^5)$ contributions and adding them to the $O(p^4)$ results
from the preceding section produces the full isospin-violating contribution
to the proton's neutral weak magnetic moment, as computed within HBChPT
(without explicit $\Delta(1232)$ fields),
\begin{eqnarray}
   G_M^{u,d}(q^2) &=& {}^{u+d}\kappa^{p-n} - {}^{u-d}\kappa^{p+n}
             - \frac{16g_A^2M_N}{(4\pi{F})^2}(M_n-M_p)
               \int_0^1{\rm d}x\,{\rm ln}\left(1-x(1-x)\frac{q^2}{m_{\pi^+}^2}
               \right) \nonumber \\
         && - (M_n-M_p)\frac{24\pi{g}_A^2m_{\pi^+}}{(4\pi{F})^2}
              \left[\frac{\mu_p+\mu_n}{\mu_N}-\frac{1}{2}
            - \frac{4}{3}\int_0^1{\rm d}x\sqrt{1-x(1-x)\frac{q^2}{m_{\pi^+}^2}} 
            \right. \nonumber \\
         && \left. + \frac{1}{6}\int_0^1{\rm d}x\left(
              1-x(1-x)\frac{q^2}{m_{\pi^+}^2}\right)^{-1/2}\right].
         \label{GMnodelta}
\end{eqnarray}
The parameters $a_6$ and $a_7$ have been re-expressed as the nucleon magnetic
moments via Eq.~(\ref{L2bit}),
\begin{eqnarray}
   a_6 &=& \frac{\mu_p-\mu_n}{4\mu_N} \approx 1.176, \\
   a_7 &=& \frac{\mu_p+\mu_n}{2\mu_N} \approx 0.440,
\end{eqnarray}
plus higher-order corrections that are not needed for the present work.
$\mu_N$ is the nuclear magneton.

\begin{figure}[tbh]
\epsfxsize=380pt \epsfbox[30 419 498 732]{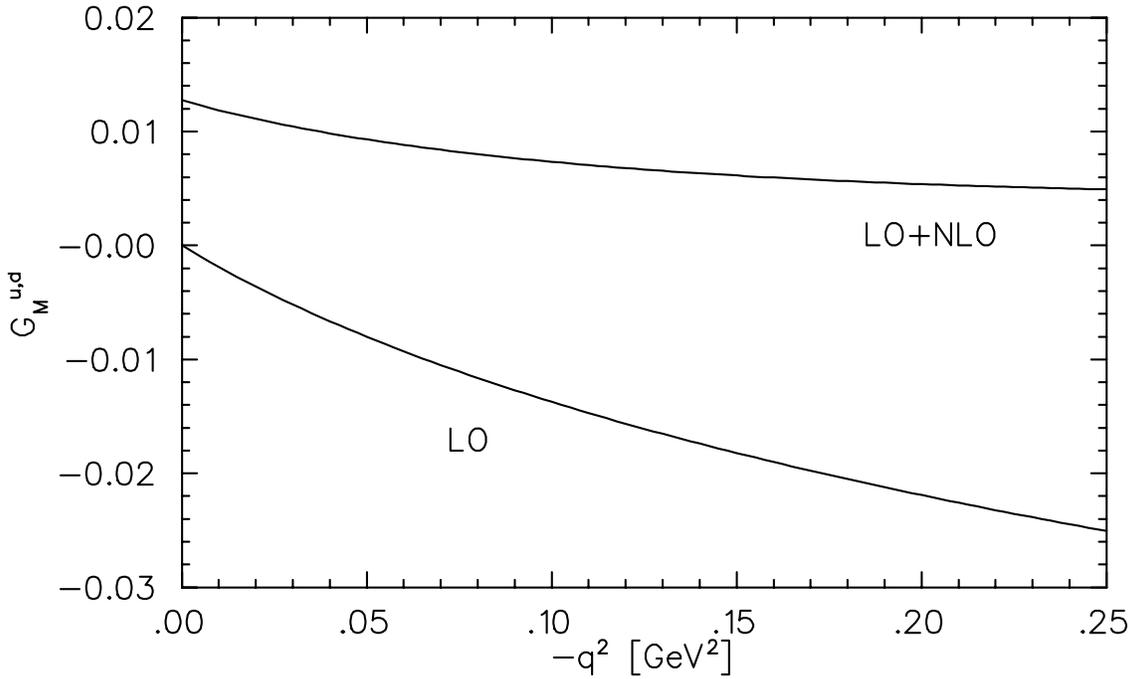}
\vspace{18pt}
\caption{\small
         The leading order (LO) and next-to-leading order (NLO) HBChPT
         results for the isospin-violating contribution to the proton's
         neutral weak magnetic moment.  For this plot, the unspecified 
         constant at LO has been chosen such that the form factor 
         vanishes at $q^2=0$.  There are no unspecified parameters in
         the NLO contribution.
         }\label{O5plot}
\end{figure}
Fig.~\ref{O5plot} contains a plot of $G_M^{u,d}(q^2)$ versus $q^2$ at leading 
order (LO) and next-to-leading order (NLO), with the only unspecified 
quantity ${}^{u+d}\kappa^{p-n} - {}^{u-d}\kappa^{p+n}$ set to zero.
The NLO effects serve to soften the $q^2$-dependence of the form factor.
Fig.~\ref{O5plot} indicates that the NLO corrections to $G_M^{u,d}(q^2)$ total
roughly 0.01 at $q^2=0$, and grow to about 0.02 near $q^2=0.1~{\rm GeV}^2$.

To determine how well the HBChPT expansion is working for this observable,
it is useful to consider the derivative of the form factor at $q^2=0$,
\begin{equation}\label{error}
   \frac{\rm d}{{\rm d}(-q^2)}G_M^{u,d}(0) = -\frac{8g_A^2M_N(M_n-M_p)}
   {3m_{\pi^+}^2(4{\pi}F)^2}\left(1-\frac{9{\pi}m_{\pi^+}}{8M_N}\right).
\end{equation}
Thus the ratio of magnitudes of the NLO/LO contributions is 
$9{\pi}m_{\pi^+}/8M_N \approx 1/2$.
Taking this as representative, the uncertainty from the neglect of NNLO
effects could be roughly half of the NLO contribution.
These uncertainties will be discussed further in section~\ref{conc}.

\vspace{3mm}
\section{\small INCLUDING THE DELTA RESONANCE}\label{delta}

Conspicuous by its absence from the preceding discussion is the $\Delta(1232)$
isomultiplet.
The effects of an infinitely-heavy $\Delta(1232)$ are accommodated within the 
numerical values of the HBChPT parameters, but the rather small
$\Delta$-$N$ mass difference observed in nature raises the possibility of
substantial corrections to the $M_\Delta \rightarrow \infty$ limit.

The incorporation of an explicit $\Delta(1232)$ field into HBChPT was
initiated by Jenkins and Manohar\cite{JenMan} and has been employed by
various authors.\cite{Delusers,HHK}
In this section, the formalism developed by Hemmert, Holstein and Kambor 
in Ref. \cite{HHK} is used to calculate
the leading $\Delta(1232)$ contributions to the isospin-violating 
${}^iF_1^j$ and ${}^iF_2^j$ form factors.

As is familiar from relativistic approaches to spin-3/2 field theory, the
Lagrangian contains a vector-spinor and one then employs a projection 
operator to isolate the spin-3/2 piece.
For example, the lowest-order propagator in ``$d$'' spacetime dimensions is
\begin{equation}
   \left(\frac{-i}{v{\cdot}k-\Delta+i\epsilon}\right)\left[g_{\mu\nu}
   -v_\mu{v}_\nu+\left(\frac{4}{d-1}\right)S_\mu{S}_\nu\right],
\end{equation}
where the incoming 4-momentum is $M_{\Delta,0}v_\mu + k_\mu$, $M_{\Delta,0}$
is the lowest-order $\Delta(1232)$ mass, and
\begin{equation}
   \Delta \equiv M_{\Delta,0} - M_0.
\end{equation}

\begin{figure}[tbh]
\epsfxsize=300pt \epsfbox[-75 200 475 650]{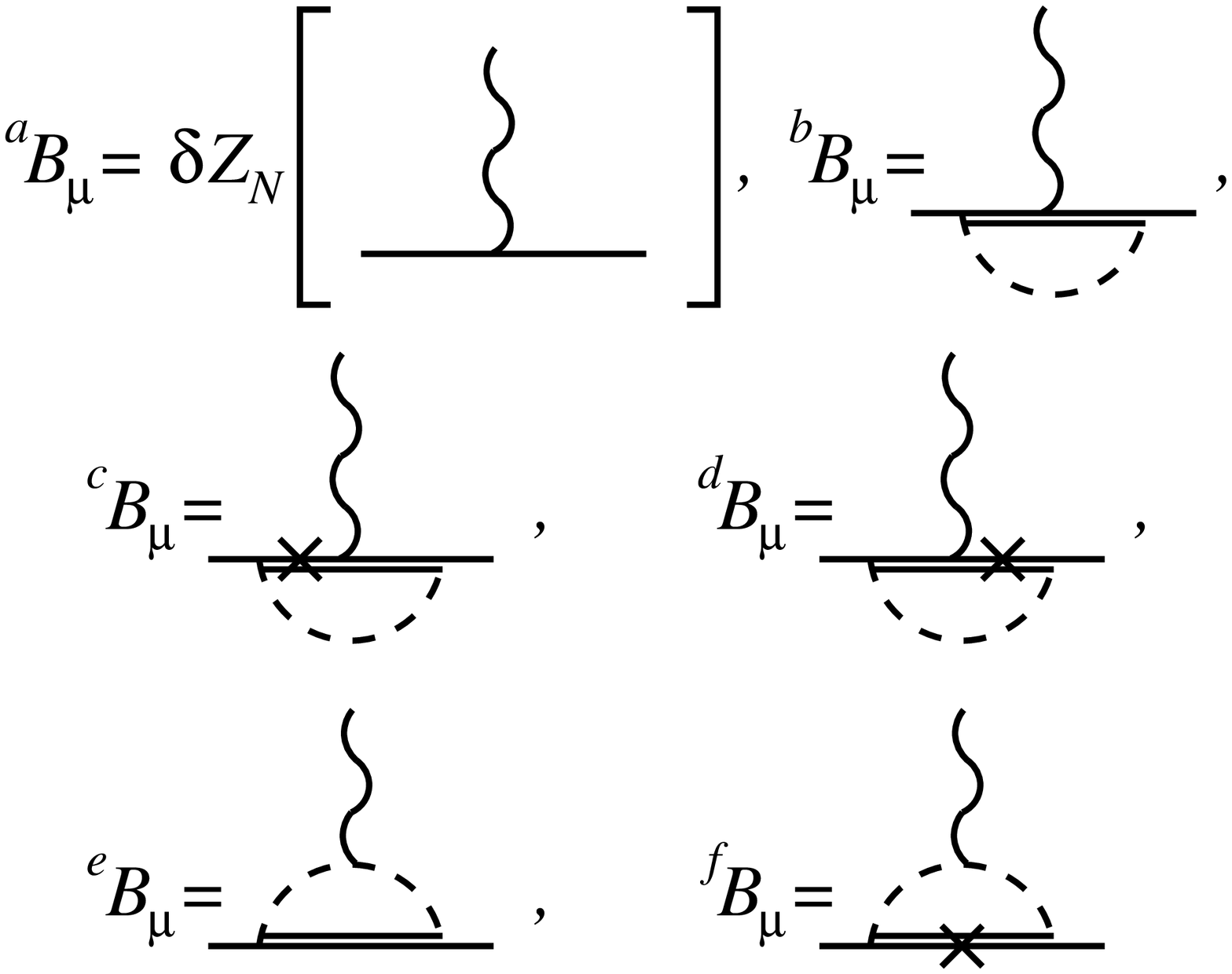}
\vspace{18pt}
\caption{\small
         Contributions of the $\Delta(1232)$ to the isospin-violating
         vector form factors of a nucleon at $O(p^4)$.
         A dashed line represents the sum over charged and neutral pions,
         and a double line represents the $\Delta(1232)$.
         A cross denotes an insertion from ${\cal L}_{\pi\Delta}^{(2)}$,
         and all other Feynman rules come from ${\cal L}_{\pi{N}}^{(1)} +
         {\cal L}_{\pi{N}\Delta}^{(1)} + {\cal L}_{\pi\pi}^{(2)}$.
         }\label{Ddiagrams}
\end{figure}
As expected, the leading contributions of the $\Delta(1232)$ to the
isospin-violating form factors appear at $O(p^4)$.
The relevant diagrams are displayed in Fig.~\ref{Ddiagrams}, and the
terms required from the leading-order Lagrangian are
\begin{eqnarray}
   \delta{\cal L}_{\pi\Delta}^{(1)} &=& -\bar{T}_i^\mu\left[iv{\cdot}D^{ij}-
   \delta^{ij}\Delta\right]T_{\mu\,j}, \\
   \delta{\cal L}_{\pi{N}\Delta}^{(1)} &=& g_{\pi{N}\Delta}\left[\bar{T}_i^\mu
   \omega^i_\mu{N} + \bar{N}\omega^{i\dagger}_\mu{T}^\mu_i\right],
\end{eqnarray}
where $T_\mu^i$ is the vector-spinor, $\omega^i_\mu$ is defined by
\begin{equation}
   \omega_\mu^i = \frac{1}{2}{\rm Tr}\left(\tau^iu_\mu\right),
\end{equation}
($\tau^i$ is a Pauli matrix in isospin space) and the covariant derivative is
\begin{equation}
   D^{ij}_\mu = \delta^{ij}(\partial_\mu+\Gamma_\mu-iv^{(s)}_\mu)
              - 2i\epsilon^{ijk}\Gamma^k_\mu.
\end{equation}
The only insertions from ${\cal L}^{(2)}$ that contribute are the 
$\Delta(1232)$ mass corrections, which arise from electromagnetic as well
as strong interaction effects.

The authors of Ref.~\cite{HHK} chose not to perform the $\Delta(1232)$ field 
transformation that would have removed ``equation of motion'' terms from 
${\cal L}^{(2)}$.  None of the calculations in this section depend
upon whether or not the transformation is performed.

\begin{figure}[tbh]
\epsfxsize=300pt \epsfbox[-75 275 475 600]{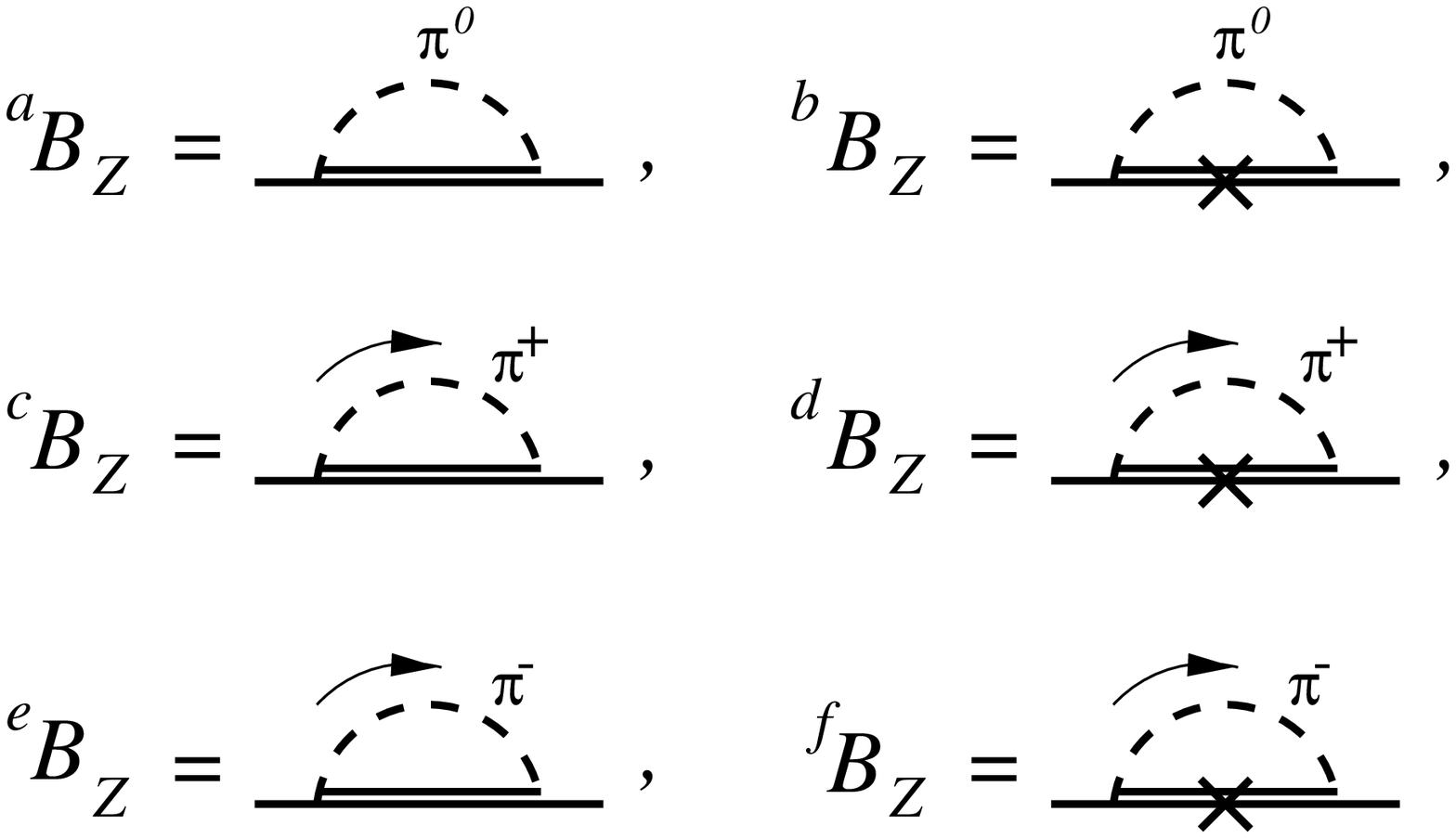}
\vspace{18pt}
\caption{\small
         Contributions of the $\Delta(1232)$ to the isospin-violating piece 
         of the nucleon's wave function renormalization constant.
         A double line represents the $\Delta(1232)$.
         A cross denotes an insertion from ${\cal L}_{\pi\Delta}^{(2)}$,
         and all other Feynman rules come from ${\cal L}_{\pi\Delta}^{(1)} +
         {\cal L}_{\pi{N}\Delta}^{(1)} + {\cal L}_{\pi\pi}^{(2)}$.
         }\label{DZdiagrams}
\end{figure}
The leading contributions of the $\Delta(1232)$ to nucleon wave function
renormalization are shown diagrammatically in Fig.~\ref{DZdiagrams}. 
When their contribution is added to the non-$\Delta(1232)$ result
of Eq.~(\ref{deltaZ}), the full isospin-violation due to wave function
renormalization is found to be
\begin{eqnarray}\label{DdeltaZ}
   Z_p - Z_n &=& \frac{M_n-M_p}{M_0}
           - 6(M_n-M_p)\frac{\pi{g}_A^2m_{\pi^+}}{(4\pi{F})^2} \nonumber \\
        && -\frac{8}{3}[4(M_n-M_p)+(M_{\Delta^0}-M_{\Delta^+})-3(M_{\Delta^-}
           -M_{\Delta^{++}})]R(m_{\pi^+}^2,0) \nonumber \\
        && - \frac{16}{3}[M_n-M_p-(M_{\Delta^0}-M_{\Delta^+})]R(m_{\pi^0}^2,0).
\end{eqnarray}
where
\begin{eqnarray}
   R(m^2,q^2) & \equiv & \frac{g_{\pi{N}\Delta}^2}{(4{\pi}F)^2}
             \left[\Delta\left(\frac{1}{\epsilon}-\gamma+{\rm ln}(4\pi)
                   -{\rm ln}\frac{m^2}{\mu^2}\right)
                   -\Delta\int_0^1{\rm d}x\,{\rm ln}\left(1-x(1-x)\frac{q^2}
                   {m^2}\right)\right. \nonumber \\
              &&   -\frac{\Delta}{3}\int_0^1{\rm d}x\left(
                   \frac{x(1-x)q^2}{\Delta^2-m^2+x(1-x)q^2}\right) \nonumber \\
              &&   -\int_0^1{\rm d}x\left(\frac{2\Delta^2-m^2+x(1-x)q^2}
                   {\sqrt{\Delta^2-m^2+x(1-x)q^2}}-\frac{x(1-x)q^2[m^2-x(1-x)
                   q^2]}{3[\Delta^2-m^2+x(1-x)q^2]^{3/2}}\right) \nonumber \\
              &&   \left.\times\,{\rm ln}\left(
                   \frac{\Delta}{\sqrt{m^2-x(1-x)q^2}}+\sqrt{\frac
                   {\Delta^2}{m^2-x(1-x)q^2}-1}\right)\right].
\end{eqnarray}

The set of $O(p^4)$ diagrams, contained in Fig.~\ref{Ddiagrams}, produces
the following contributions to the isospin-violating form factors:
\begin{eqnarray}
   \delta\left[{}^{u+d}F_1^{p-n}(q^2)\right] &=& 
   \delta\left[{}^{u+d}F_2^{p-n}(q^2)\right] = 0 \\
   \delta\left[{}^{u-d}F_1^{p+n}(q^2)\right] &=& 
       -\frac{16}{3}\left[2(M_n-M_p)-(M_{\Delta^0}-M_{\Delta^+})-3(M_{\Delta^-}
       -M_{\Delta^{++}})\right] \nonumber \\
       && \times\,[R(m_{\pi^+}^2,q^2)-R(m_{\pi^+}^2,0)], \\
   \delta\left[{}^{u-d}F_2^{p+n}(q^2)\right] &=& 
       \frac{8g_{\pi{N}\Delta}^2}{9(4{\pi}F)^2}M_N\left[2(M_n-M_p)
       -(M_{\Delta^0}-M_{\Delta^+})-3(M_{\Delta^-}-M_{\Delta^{++}})\right]
       \nonumber \\
       && \times\,
       \left[{\rm const}~+\int_0^1{\rm d}x\,{\rm ln}\left(1-x(1-x)\frac{q^2}
       {m_{\pi^+}^2}\right)\right. \nonumber \\
       && +2\int_0^1{\rm d}x\frac{\Delta}{\sqrt{\Delta^2-m_{\pi^+}^2
       +x(1-x)q^2}} \nonumber \\
       && \left.\times\,{\rm ln}\left(\frac{\Delta}{\sqrt{m_{\pi^+}^2
       -x(1-x)q^2}}+\sqrt{\frac{\Delta^2}{m_{\pi^+}^2-x(1-x)q^2}-1}\right)
       \right].
\end{eqnarray}
Notice that the total contribution made by the $\Delta(1232)$ loop graphs
to each of the ``$u+d$'' form factors exactly vanishes.
For the isovector (``$u-d$'') current, the various contributions to 
${}^{u-d}F_1^{p+n}$
add in such a way that all divergences cancel and the familiar constraint, 
${}^{u-d}F_1^{p+n}(0)=0$ is satisfied.
As expected, the contribution to ${}^{u-d}F_2^{p+n}$ is not finite, 
and the full Lagrangian contains an $O(p^4)$ counterterm which absorbs this 
divergence.

\begin{figure}[tbh]
\epsfxsize=380pt \epsfbox[30 419 498 732]{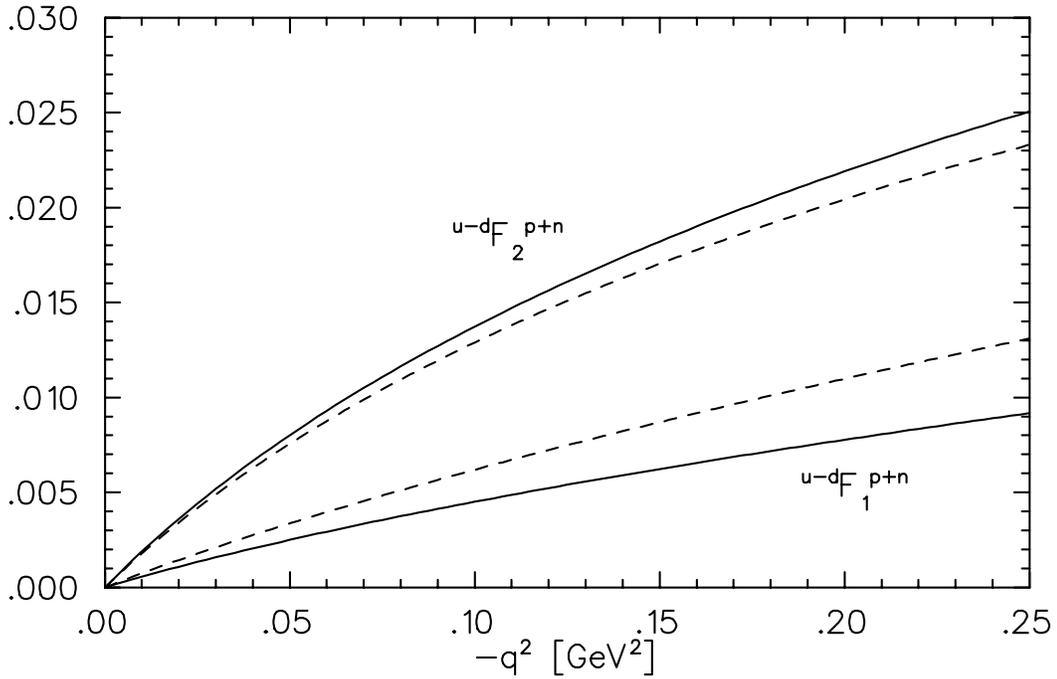}
\vspace{18pt}
\caption{\small
         Leading-order HBChPT results for the isospin-violating isovector
         form factors without (solid lines) and with (dashed lines) the
         explicit $\Delta(1232)$ isomultiplet.  
         For this plot, the unspecified constant in ${}^{u-d}F_2^{p+n}$
         has been chosen such that the form factor vanishes at $q^2=0$.
         }\label{Dplot}
\end{figure}
Plotted in Fig.~\ref{Dplot} are the leading results, i.e. the $O(p^4)$ results,
for the two ``$u-d$'' form factors with and without the explicit 
$\Delta(1232)$ contribution, and with the unspecified constant subtracted
from ${}^{u-d}F_2^{p+n}$.
The effects of the $\Delta(1232)$ are significantly smaller than 
the NLO corrections of Fig.~\ref{O5plot}.
The value $g_{\pi{N}\Delta} = 1.05$, recommended in Ref.~\cite{HHK},
has been used along with the following mass differences:
\begin{eqnarray}
   \Delta &=& 0.293~{\rm GeV}, \\
   M_{\Delta^0}-M_{\Delta^+} &=&
   \frac{1}{3}(M_{\Delta^-}-M_{\Delta^{++}}) = .0013~{\rm GeV}.
\end{eqnarray}
There are sizable experimental uncertainties on these inputs,
but no reasonable choices can make the $\Delta(1232)$ contribution
to $G_M^{u,d}(q^2)$ grow larger than the NLO correction of Fig.~\ref{O5plot}.

\vspace{3mm}
\section{\small DISCUSSION}\label{conc}

The proton's neutral weak magnetic form factor $G_M^{p,Z}$ is of great 
interest, both experimentally and theoretically, because it is sensitive
to strangeness within the nucleon.
Experiments actually measure the sum of strangeness and isospin-violating
contributions (labeled $G_M^s$ and $G_M^{u,d}$ respectively), 
so it is advantageous to understand the effects of
isospin violation as thoroughly as possible.

In this work, heavy baryon chiral perturbation theory (HBChPT) has been
used to study isospin violation in the absence of strange quarks.
HBChPT is an expansion in momenta that are small compared to the chiral
scale, $\Lambda_\chi \sim m_\rho \sim M_N \sim 4{\pi}F_\pi$.  Virtual
photons can also been included according to the usual $\alpha_{\rm QED}$
expansion.

First, $G_M^{u,d}(q^2)$ was calculated at leading order (LO) and 
next-to-leading 
order (NLO) for a single nucleon surrounded by a cloud of pions and photons.
The photon cloud was found to contain no momentum dependence at this order,
in contrast to the pion cloud.
The ratio of NLO/LO contributions was used as an estimator of the systematic
uncertainty of the HBChPT expansion.  Then the
contribution of the $\Delta(1232)$ isomultiplet was evaluated and found to
be negligible in comparison with the systematic uncertainty.

The final expression is given by Eq.~(\ref{GMnodelta}) in terms of a single
unspecified parameter, ${}^{u+d}\kappa^{p-n}-{}^{u-d}\kappa^{p+n}$. 
This parameter contains physics of two types: a low-energy contribution 
from the pion cloud, and a higher-energy contribution which is unconstrained
by HBChPT.

The pion-cloud contribution to the unspecified parameter is easily determined
by redoing the HBChPT calculation with a momentum cutoff, $\lambda$, instead of
dimensional regularization.  
This cutoff represents the separation scale between the ``low-energy'' and
``higher-energy'' regions.  In principle, the choice of $\lambda$ does not 
affect $G_M^{u,d}(q^2)$ since only the {\it sum\/} of low-energy and
higher-energy pieces is relevant, but in HBChPT the higher-energy piece is
undetermined so it is preferable to make a physical choice for $\lambda$.

Clearly $\lambda$ cannot be larger than $\Lambda_\chi$, since HBChPT fails 
above this scale.  In a series of recent papers\cite{DHB}, 
Donoghue, Holstein and Borasoy have argued for an HBChPT cutoff that is not 
too far above 1 fm${}^{-1} \approx$ 200 MeV, corresponding to the measured 
size of a baryon.  Above this approximate scale the substructure of a
nucleon can begin to be relevant, but is not accurately represented in HBChPT.

The only divergence in the non-$\Delta(1232)$ piece of $G_M^{u,d}(q^2)$ comes
from a single integral which appears in diagrams ${}^fA_\mu$ and ${}^gA_\mu$
of Figure \ref{O4diagrams}.  When dimensional 
regularization is replaced by a simple cutoff for the momentum integral,
the following relationship is obtained:
\begin{equation}
   \left(
   \frac{2}{4-d}-\gamma+{\rm ln}(4\pi)-{\rm ln}\frac{m_{\pi^+}^2}{\mu^2}
   \right)
   ~~~\rightarrow~~~
   \left({\rm ln}\frac{\lambda^2}{m_{\pi^+}^2} - \frac{3}{2}\right).
\end{equation}
Choosing $\lambda =$ 400 MeV, the low-energy contribution of the pion cloud
is found to be
\begin{equation}
   \left[{}^{u+d}\kappa^{p-n}-{}^{u-d}\kappa^{p+n}\right]_{\rm pion~cloud} 
   = 0.014.
\end{equation}
Notice that the dependence on $\lambda$ is only logarithmic, so the
result is not overly sensitive to the chosen numerical value of the cutoff.
The $\Delta(1232)$ contribution also contains a logarithmic dependence on
$\lambda$, but the resulting pion-cloud contribution is negligible in 
comparison to the uncertainties coming from the HBChPT expansion (recall
section~\ref{delta}).

The remaining contribution to the $G_M^{u,d}$ counterterm is the
``higher-energy'' contribution.  It is unspecified in HBChPT by 
definition, and a precise numerical prediction is therefore beyond the scope 
of this work.
If the HBChPT expansion is to be well-behaved, then the higher-energy
contributions must respect the established power counting.
Perhaps the most obvious examples of higher-energy physics are 
the $\rho$ and $\omega$ vector mesons.
Simple power counting estimates for the tree-level vector meson
dominance diagrams indicate that their leading contribution resembles
\begin{equation}
   \left[G_M^{u,d}\right]_{\rho,\omega} \propto
   \frac{M_N(M_n-M_p)}{m_{\rho,\omega}^2},
\end{equation}
which is $O(p^4)$, as required by HBChPT.

\begin{figure}[tbh]
\epsfxsize=380pt \epsfbox[30 419 498 732]{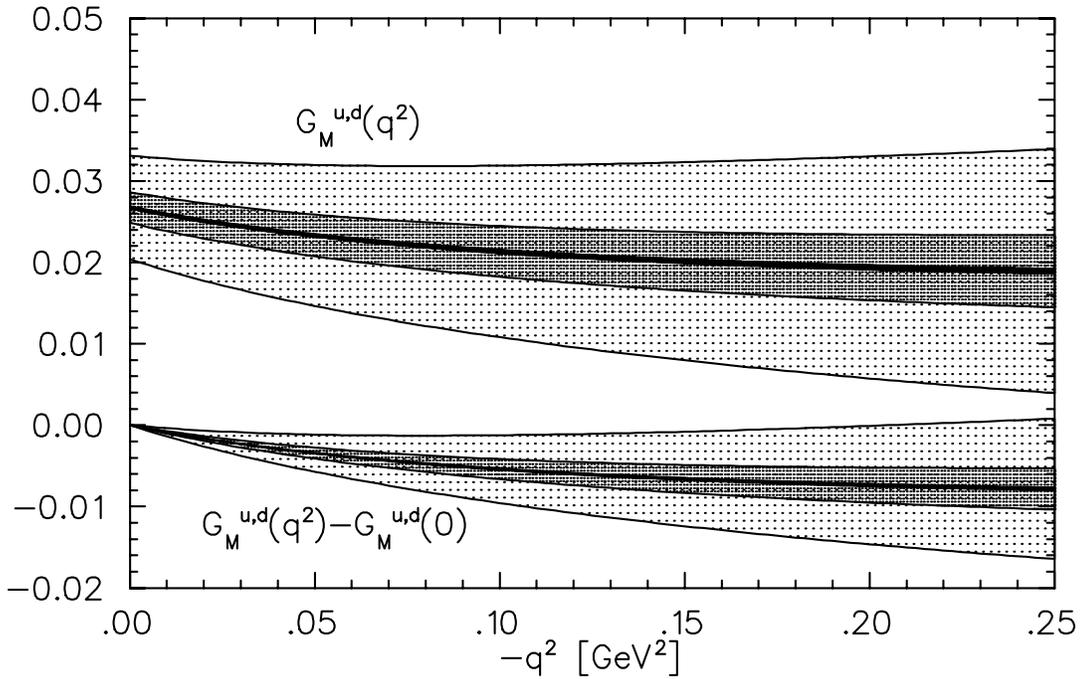}
\vspace{18pt}
\caption{\small
         The HBChPT prediction for the isospin-violating pion-cloud 
         contribution to the proton's neutral weak magnetic form factor.
         The two solid lines represent the central value with and
         without a subtraction at $q^2=0$, and a pair of
         uncertainty estimates are provided for each solid line
         according to Eq.~(\protect\ref{hatch}).
         The logarithmic divergence is cut off at 400 MeV,
         as discussed in the text.  
         }\label{finalfig}
\end{figure}
The main conclusion of this work is summarized by Fig.~\ref{finalfig},
which shows the full contribution of the pion cloud to $G_M^{u,d}(q^2)$, 
for $0 < -q^2 < 0.25~{\rm GeV}^2$, up to next-to-leading order in the HBChPT
expansion.  To aid a discussion of $q^2$-dependence, the same quantity is
shown after $G_M^{u,d}(0)$ has been subtracted.  
In each case, a pair of uncertainty bands is shown, representing two 
estimates of the error associated with the neglect of NNLO contributions,
\begin{equation}\label{hatch}
   |{\rm NNLO}| \sim \left\{ \begin{array}{l}
                  \frac{1}{2}|{\rm NLO}|~,~~{\rm wide~band} \\
                  \frac{m_\pi}{M_N}|{\rm NLO}|~,~~{\rm narrow~band} 
                  \end{array} \right.~.
\end{equation}
The wide-band error estimate purports that the ratio of 1/2, taken from
$|{\rm NLO}|/|{\rm LO}|$ in Eq.~(\ref{error}) for the derivative of 
$G_M^{u,d}$, might be a reasonable indicator of the NNLO uncertainty.
The narrow-band error estimate employs a generic HBChPT expansion parameter.

Fig.~\ref{finalfig} indicates that the $q^2$-dependence, as determined by
$G_M^{u,d}(q^2)-G_M^{u,d}(0)$, occurs on the scale of a few times 0.001, 
but is typically less than 0.01.
This result is easily understood: HBChPT demands that the momentum dependence
of this isospin-violating form factor is proportional to 
$(M_n-M_p)/M_N \approx 0.001$.  Recall that no contributions from 
$m_{\pi^+} - m_{\pi^0}$ or any of the HBChPT parameters were found.

However, Fig.~\ref{finalfig} also indicates that the contributions of the 
pion cloud to $G_M^{u,d}(0)$ are on the scale of a few times 0.01.
The origin of isospin violation is still solely $(M_n-M_p)/M_N \approx 0.001$, 
but the numerical coefficients are larger than those for the $q^2$-dependence.

It is interesting to compare the results of Fig.~\ref{finalfig} to the
findings of other authors.
Dmitra\v{s}inovi\'{c} and Pollock have used a nonrelativistic constituent
quark model to find\cite{DmiPol}
\begin{equation}
   \left[{}^{u+d}G_M^{p-n}(0)\right]_{\rm DmiP} = 
   \left[{}^{u-d}G_M^{p+n}(0)\right]_{\rm DmiP} \approx 0.008
   ~~~\Rightarrow~~~
   \left[G_M^{u,d}(0)\right]_{\rm DmiP} = 0.
\end{equation}
A vanishing total result at $q^2=0$ is also obtained by Miller, who has 
studied a 
family of three nonrelativistic constituent quark models.\cite{Miller}
As well, his work suggests that the $q^2$-dependence is very mild:
\begin{equation}
   \left[G_M^{u,d}(0) -
         G_M^{u,d}(-0.25~{\rm GeV}^2)\right]_{\rm Mil} < 0.001.
\end{equation}
Capstick and Robson have work in progress that employs a relativized 
constituent quark model.\cite{CapRob}
Using a light-cone meson-baryon fluctuation model, Ma has reported the
following allowed range\cite{Ma}:
\begin{equation}
   \left[G_M^{u,d}(0)\right]_{\rm Ma} = 0.006 \rightarrow 0.088.
\end{equation}

In light of the uncertainties assigned to HBChPT, there is no essential
disagreement between the present work and any of these models.  
Certainly the tendency of
nonrelativistic quark models to prefer $G_M^{u,d}(0) = 0$ is not
obtained from the pion cloud in HBChPT, but recall that the effect of 
higher-energy physics remains unspecified in HBChPT.

In conclusion, the effects of isospin violation on the proton's neutral
weak magnetic form factor have been studied up to next-to-leading order
in heavy baryon chiral perturbation theory.  The momentum dependence
contains no free parameters, and comes solely from the neutron-proton
mass difference despite the large number of parameters in the Lagrangian.  
Normalization of the isospin-violating contribution
at $q^2=0$ is not specified by chiral symmetry, but the pion-cloud
effects can be extracted and their contribution is roughly 0.02
nuclear magnetons.

\vspace{3mm}
\section*{\small ACKNOWLEDGMENTS}

We are grateful to Jos\'e Goity for a critical reading of the manuscript.
This research was supported in part by the Natural Sciences and Engineering
Research Council of Canada.
R.L. also acknowledges support from the U.S. Department of Energy,
contract DE-AC05-84ER40150.


\end{document}